\documentstyle{article}
\begin{document}
\noindent
SAGA-HE-90-95

\noindent
August 1995

\bigskip

\bigskip

\bigskip

\centerline{\bf Volume, Coulomb, and volume-symmetry coefficients}

\centerline{\bf of nucleus incompressibility in the relativistic mean field
theory}

\centerline{\bf with the excluded volume effects}

\bigskip

\bigskip

\centerline{\bf{H. Kouno, T. Mitsumori, K. Koide, N. Noda, A. Hasegawa}}

\centerline{Department of Physics, Saga University, Saga 840, Japan}

\centerline{\bf{and}}

\centerline{\bf{ M. Nakano}}

\centerline{University of Occupational and Environmental Health, Kitakyushu
807, Japan}

\bigskip

\bigskip


\bigskip

\bigskip

\centerline{\bf ABSTRACT}

\noindent
The relation among the volume coefficient $K$(=incompressibility of the nuclear
matter), the Coulomb coefficient $K_c$, and the volume-symmetry coefficient
$K_{vs}$ of the nucleus incompressibility are studied in the framework of the
relativistic mean field theory with the excluded volume effects of the
nucleons, under the assumption of the scaling model.
It is found that $K= 300\pm 50$MeV is necessary to account for the empirical
values of $K$, $K_c$, and $K_{vs}$, simultaneously, as is in the case of the
point-like nucleons. The result is independent on the detail descriptions of
the potential of the $\sigma$-meson self-interaction and is almost independent
on the excluded volume of the nucleons.

\vfill\eject

To determine the incompressibility $K$ of nuclear matter from the giant
monopole resonance (GMR) data, one may use the leptodermous expansion [1] of
nucleus incompressibility $K(A,Z)$ as follows.
$$  K(A,Z)=K+K_{sf}A^{-1/3}+K_{vs}I^2+K_cZ^2A^{-4/3}+\cdot \cdot
\cdot~~~~~;~~~I=1-2Z/A, \eqno{(1)} $$
where the coefficients, $K$, $K_{sf}$, $K_{vs}$ and $K_c$ are volume
coefficient (incompressibility of nuclear matter), surface term coefficient,
volume-symmetry coefficient and Coulomb coefficient, respectively.
The higher terms is omitted in eq. (1).
Although there is uncertainty in the determination of these coefficients by
using the present data, Pearson [2] pointed out that there is a strong
correlation between $K$ and $K_c$.
(See table 1.)
Similar observations are done by Shlomo and Youngblood [3].


\centerline{$\underline{~~~~~~~}$}

\centerline{Table 1}

\centerline{$\underline{~~~~~~~}$}


According to this context, Rudaz et al. [4] studied the relation between
incompressibility and the skewness coefficient by using the generalized version
of the relativistic Hartree approximation [5].
The compressional and the surface properties are studied by Von-Eiff et al.
[6-8] in the framework of the relativistic mean field theory (RMF) of the
$\sigma$-$\omega$-$\rho$ model with the nonlinear $\sigma$ terms.
They found that low incompressibility ($K\approx 200$MeV) and a large effective
nucleon mass $M^*$ at the normal density ($0.70\leq M^*/M\leq 0.75$) are
favorable for the nuclear surface properties [8], where $M$ is a free nucleon
mass.
On the other hand, using the same model, Bodmer and Price [9] found that the
experimental spin-orbit splitting in light nuclei supports $M^*\approx 0.60M$.
Furthermore, the result of the generator coordinate calculations for
breathing-mode GMR by Stoitsov, Ring and Sharma [10] suggests $K\approx
300$MeV.
It seems that there are two suggestions, large ($>0.7M$) or small ($<0.7M$)
$M^*$, and there are also two suggestions, large ($\sim 300$MeV) or small
($\sim 200$MeV) $K$.

In refs. [11,12,13], we studied the relation between $K$ and $K_c$ in detail,
by using RMF with the nonlinear $\sigma$ terms [14], by using RMF with the
nonlinear $\sigma$ and $\omega$ terms [15], and by using RMF with the nonlinear
$\sigma$ term and the excluded volume effects (EVE) of nucleons[16].
One of our finding in those studies is that, under the assumption of the
scaling model [1], $K=300\pm 50$MeV is favorable to account for $K$, $K_c$ and
$K_{vs}$, simultaneously, in any case of three models.
The conclusion seemed to be common in the RMF, although, in the three analyses,
we assume the quartic-cubic potential as the $\sigma$ meson self-interaction.
In recent paper[17], we have reexamined the conclusion by using RMF with the
nonlinear $\sigma$ terms, and by using the one with the nonlinear $\sigma$ and
$\omega$ terms, in more general way in which the result does not depend on the
detail description of the $\sigma$ meson self-interaction $U(\phi )$.
It is found that the conclusion $K=300\pm 50$MeV is $necessary$ for $any$ type
of $U(\phi )$ in both cases.
It is also found that the result hardly depends on the strength of the $\omega$
meson self-interaction.
It seems that this conclusion is not drastically changed in the use of any type
of RMF and the scaling model.
In this paper, we reexamine the conclusion by using the RMF with the nonlinear
$\sigma$ terms and EVE of the nucleons [16, 13], in the general way in which
the result does not depend on $U(\phi )$ as in ref. [17].
The reason why we restrict our discussions to $K$, $K_c$ and $K_{vs}$ is that
the general discussions, which are independent of the detail of the model (
e.g., types of the interactions, values of the parameters in the Lagrangian,
etc.) are possible to a considerable extent, since, as is shown below, these
quantities are almost analytically estimated by using the result for the
nuclear matter, if we assume the scaling model [1].
We also remark that our discussions do not depend on the $\sigma$ meson mass,
since $K$, $K_c$, and $K_{vs}$ are able to be calculated by the result for the
nuclear matter.

We use RMF with EVE of nucleons [16, 13].
For the Lagrangian density, we use the $\sigma$-$\omega$-$\rho$ model with the
nonlinear $\sigma$ terms.
The Lagrangian density consists of four fields, the nucleon $\psi$, the scalar
$\sigma$-meson $\phi$, the vector $\omega$-meson $V_\mu$, and the
vector-isovector $\rho$ meson ${\bf b}_\mu$, i.e.,
$$ L_{N\sigma\omega\rho} =\bar{\psi}(i\gamma_\mu\partial^\mu -M)\psi$$
$$
+{1\over{2}}\partial_\mu\phi\partial^\mu\phi -{1\over{2}}m_s^2\phi^2
-{1\over{4}}F_{\mu\nu} F^{\mu\nu} +{1\over{2}}m_v^2V_\mu V^\mu
-{1\over{4}}{\bf{B}}_{\mu\nu}\cdot{\bf{B}}^{\mu\nu}+{1\over{2}}m_\rho^2{\bf{b}}_\mu\cdot{\bf{b}}^\mu$$
$$ +g_s\bar{\psi}\psi\phi-g_v\bar{\psi}\gamma_\mu \psi V^\mu
-g_\rho\bar{\psi}\gamma_\mu{\tau\over{2}}\cdot{\bf {b}}^\mu\psi-U(\phi )~~~;$$
$$F_{\mu\nu}=\partial_\mu V_\nu -\partial_\nu
V_\mu,~~~~~{\bf{B}}_{\mu\nu}=\partial_\mu {\bf{b}}_\nu -\partial_\nu
{\bf{b}}_\mu-g_\rho {\bf b}_\mu\times {\bf b}_\nu , \eqno{(2)} $$
where $m_s$, $m_v$, $m_\rho$, $g_s$, $g_v$ and $g_\rho$ are $\sigma$-meson
mass, $\omega$-meson mass, $\rho$-meson mass, $\sigma$-nucleon coupling,
$\omega$-nucleon coupling, and $\rho$-nucleon coupling, respectively.
The $U(\phi )$ is a nonlinear self-interaction potential of $\sigma$ meson
field $\phi$.
For example, in ref. [13], we have used the quartic-cubic terms of $\phi$ as in
ref. [14], i.e.,
$$
U(\phi ) ={{1}\over{3}}b\phi^3+{{1}\over{4}}c\phi^4, \eqno{(3)} $$
where $b$ and $c$ are the constant parameters which are determined
phenomenologically.
However, in this paper, we do not give an explicit expression of $U(\phi)$ and
discuss the problem in more general way, without any assumption of $U(\phi )$,
as in ref. [17].

In RMF of the point-like nucleons, the baryon density is given by
$$ \rho^{pt}={\lambda\over{3\pi^2}}k_F^3,    \eqno{(4)} $$
where $k_F$ is Fermi momentum and $\lambda =2$ in the nuclear matter.
In the model with the EVE [16, 13], the volume $V$ for the $N$ body system of
nucleons in configurational space is reduced to the effective one, $V-NV_n$,
where $V_n$ is the excluded volume of a nucleon.
According to this modification for the volume, the baryon density $\rho$ is
given by
$$ \rho={{\tilde{\rho}}\over{1+V_n{\tilde{\rho}}}}, \eqno{(5)} $$
where ${\tilde{\rho}}$ has the same expression as $\rho_{pt}$ for the given
$k_F$.
In a similar way, the scalar density is given by
$$ \rho_s={{{\tilde{\rho}}_s}\over{1+V_n{\tilde{\rho}}}} \eqno{(6)} $$
where ${\tilde{\rho}}_s$ has the same expression as the scalar density of the
system of the point-like nucleons, and is given by
$$
{\tilde{\rho}}_s={\lambda\over{2\pi^2}}M^*[k_F\sqrt{k_F^2+M^{*2}}-M^{*2}\ln{({{k_F+\sqrt{k_F^2+M^{*2}}}\over{M^*}})}].    \eqno{(7)} $$
For the detail description of this model, see refs. [16, 13].

In the scaling model [1], $K$, $K_c$ and $K_{vs}$ in eq. (1), are given by
$$  K=9\rho_0^2{{\partial^2E_b}\over{\partial \rho^2}}\vert_{\rho=\rho_0},
\eqno{(8)} $$
$$  K_c=-{{3q_{el}^2}\over{5R_0}}\biggl( {9K'\over{K}}+8 \biggr), \eqno{(9)} $$
$$  K_{vs}=K_{sym}-L\biggl( 9{K'\over{K}}+6 \biggr), \eqno{(10)} $$
where $\rho$, $\rho_0$, $E_b$ and $q_{el}$ are the baryon density, the normal
baryon density, the binding energy per nucleon, and the electric charge of
proton, respectively, and $R_0=[3/(4\pi\rho_0)]^{1/3}$,
$$ K'=3\rho_0^3{{d^3 E_b}\over{d \rho^3}}\vert_{\rho=\rho_0}, \eqno{(11)} $$
$$  L=3\rho_{0}{{d J}\over{d \rho}}\vert_{\rho =\rho_{0}},
{}~~~~K_{sym}=9\rho_{0}^2{{d^2 J}\over{d \rho^2}}\vert_{\rho =\rho_{0}}:
{}~~~~J={1\over{2}}\rho^2{{\partial^2E_b}\over{\partial\rho_3^2}}\vert_{\rho_3=0}.  \eqno{(12)}  $$
The quantity such as $K'$ is sometimes called "skewness".

In RMF with EVE, $L$ and $K_{sym}$ are given by [7]
$$  L={{3\rho}\over{8}}\alpha_\rho^2+{\rho\over{2}}\biggl(
{{2k_Fk^{\prime}_F}\over{E_F^*}}-{{k_F^2E_F^{*\prime}}\over{E_F^{*2}}}\biggr)~~~~~~(\rho =\rho_0), \eqno{(13)} $$
and
$$  K_{sym}={3\over{2}}\rho^2\biggl(
{{2k_F^{\prime2}}\over{E_F^*}}+{{2k_Fk_F^{\prime\prime}}\over{E_F^*}}-{{4k_Fk_F^{\prime}E_F^{*\prime}}\over{E_F^{*2}}}+{{2k_F^2E_F^{*\prime 2}}\over{E_F^{*3}}}-{{k_F^2E_F^{*\prime\prime}}\over{E_F^{*2}}} \biggr)~~~~~(\rho =\rho_0),    \eqno{(14)} $$
where $\alpha_\rho =g_\rho /m_\rho$,
$$
k'_F={{dk_F}\over{d\rho}}=h{{k_F}\over{3\rho}},~~~~~k_F^{\prime\prime}={{d^2k_F}\over{d\rho^2}}=2h^2(V_n-{1\over{3\rho}}){{k_F}\over{3\rho}},~~~~~h=1+V_n{\tilde{\rho}}, \eqno{(15)} $$
$$    E_F^*=\sqrt{k_F^2+M^{*2}}, ~~~~~E_F^{*\prime}={{dE_F^*}\over{d\rho}},
{}~~~~~E_F^{*\prime\prime}={{d^2E_F^*}\over{d\rho^2}}.   \eqno{(16)} $$
We remark that equations for the point-like nucleon [17] are given, if we put
$V_n=0$ in eq. (15).

At $\rho =\rho_0$, $E_F^{*\prime}$ and $E_F^{*\prime\prime}$ are related to $K$
and $K'$ in the following relations, respectively.
$$
E_F^{*\prime}={{k_F}\over{E_F^*}}k_F^{\prime}+{{M^*}\over{E_F^*}}M^{*\prime},
\eqno{(17)} $$
$$ M^{*\prime}={{dM^*}\over{d\rho}}=
[{{K}\over{9\rho}}-\alpha_v^2-h{{k_F}\over{E_F^*}}{{dk_F}\over{d\rho}}]/(h{{M^*}\over{E_F^*}}-V_n{\tilde{\rho}}_s),~~~~~\alpha_v={{g_v}\over{m_v}}, \eqno{(18)}   $$
$$    E_F^{*\prime\prime}
=[{{K+K'}\over{3\rho^2}}-V_n(h^2E_F^{*\prime}-{\tilde{\rho}}_s^{\prime}M^{*\prime}-t{\tilde{\rho}}_s)]/(h-{{E_F^*}\over{M^*}}V_n{\tilde{\rho}}_s), \eqno{(19)} $$
$$   t=(E_F^{*\prime 2}-k_F^{\prime 2}-k_Fk_F^{\prime\prime}-M^{*\prime
2})/M^*, \eqno{(20)} $$
$$
{\tilde{\rho}}_s'={{d{\tilde{\rho}}_s}\over{d\rho}}=h^2{{M^*}\over{E_F^*}}+{{\lambda M^{*\prime}}\over{2\pi^2}}[k_FE_F^*+{{2k_FM^{*2}}\over{E_F^*}}-3M^{*2}\ln{({{k_F+E_F^*}\over{M^*}})}], \eqno{(21)} $$
Furthermore, at $\rho =\rho_0$, $C_v$ and $C_\rho$ are also related to $M^*$ as
follows.
$$ \alpha_v^2={{1}\over{\rho}}\biggl( M-a_1-E_F^*-V_nP(E_F^*,M^*)\biggr) ,
\eqno{(22)} $$
[13] and
$$  \alpha_\rho^2 ={{8}\over{\rho}}\biggl(
a_4-{{k_{F}^2}\over{6E_{F}^*}}\biggr), \eqno{(23)} $$
[18] where $P(E_F^*,M^*)$ is given by replacing $M$ by $M^*$ in the pressure
$P(\sqrt{k_F^2+M^2},M)$ of the free nucleon system, and $a_1$ and $a_4$ are the
binding energy and the symmetry energy at $\rho =\rho_0$, respectively.
We remark that eqs. (9), (10), and (13)$\sim$(23) have no explicit dependence
on $U(\phi )$.

{}From eq. (9), $K'$ are determined, if $\rho_0$, $K$ and $K_c$ are given.
Therefore, from the eqs. (9), (10), and (13)$\sim$(23), it is seen that
$K_{vs}$ is determined, if $\rho_0$, $a_1$, $a_4$, $K$, $K_c$, $V_n$ and $M^*$
are given, without giving the detail descriptions for $U(\phi )$.
Using these equations, we calculate $K_{vs}$.
In the calculations, we put $\rho_0=0.15$fm$^{-3}$, $a_1=E_b(\rho_0)=15.75$MeV
and $a_4=J(\rho_0)=30.0$MeV.
We restrict $R_n(=[3V_n/4\pi ]^{1/3})$ in the region of 0$\sim$0.8fm, because
larger value of $R_n(~^>_\sim 0.9$fm), which is close to the average nucleon
spacing $R_0(\sim 1.17$fm) in normal nuclear matter, may cause divergences in
calculations.
For $K$ and $K_c$, we use the values in table 1.
We assume that $M^*=0.5M\sim 0.93M$, the phenomenologically acceptable values.
( The upper bound for $M^*$ is gotten, if we put $C_v=0$ and $k_F\sim
1.4$fm$^{-1}$ in eq. (21). We remark that $k_F$ is larger in RMF with EVE than
in the case of the point-like nucleons. )
In fig. 1, we show $K_{vs}$ as a function of $M^*$ for two sets of $K$ and
$K_c$ in table 1, at $R_n=0.8$fm and at $R_n=0.6$fm, comparing them with the
result in the case of the point-like nucleons.
In the fig. 1(a), $K_{vs}$ decreases as $M^*$ increases in any case of $R_n$.
In that case (in the case of $K=300$MeV and $K_c=-3.990$MeV),
$K_{vs}=-247(-292,-298)\sim 176(-53,-115)$MeV for $R_n=0.8(0.6,0)$fm.
These values are in good agreement with the corresponding empirical values in
table 1.
We also remark that the uncertainty of $K_{vs}$ in changing $M^*$ is not so
larger than the empirical error bar of $K_{vs}$ in table 1.
In the case of $K=350$MeV and $K_c=-7.274$MeV (fig. 1(b)), $K_{vs}$ shows more
complicated behaviors in changing $M^*$.
However, the uncertainty of $K_{vs}$ is comparable to the magnitude of the
empirical error bar of $K_{vs}$ in table 1.
In this case, $K_{vs}=-652(-673, -671)\sim -484(-610,-632)$MeV for
$R_n=$0.8(0.6,0)fm.
These values are somewhat smaller than the empirical ones.
In table 2, we summarize the range of the calculated $K_{vs}$ for each set of
$K$ and $K_c$ in table 2.
Comparing the table 1 and table 2, we see that $K=300\pm 50$MeV is necessary to
account for $K$, $K_c$ and $K_{vs}$, simultaneously in any case of $R_n$, as in
the case of the point-like nucleon ($R_n=0$) [17].
To say more exactly, in many cases of $K $ and $M^*$, the repulsive effect of
EVE tends to make $K_{vs}$ larger, on the contrary to the attractive effects of
the vector meson self-interaction (VSI) [12, 17], in which $K_{vs}$ tends to
becomes smaller by introducing VSI.
However, these effects are not so large enough to change the conclusion.


\centerline{$\underline{~~~~~~~~~~~~~~~~~~~~~}$}

\centerline{Fig. 1(a),(b), Table 2(a),(b),(c)}

\centerline{$\underline{~~~~~~~~~~~~~~~~~~~~~}$}


We remark the following three points.

(1) The results are independent of the form of $U(\phi )$, since eqs. (9),
(10), and (13)$\sim$(23) are required for any type of $U(\phi )$.

(2) The question, whether there are coupling parameters, which reproduce the
set of $K$ and $K_c$ in table 1, or not, is still open, and the answer for the
question depends on the detail descriptions of $U(\phi )$.
For example, if we use the quartic-cubic potential (3),
we could not find the coupling parameters, which reproduce $K=200$MeV and
$K_c=2.577$MeV at any $R_n$ [13].
Also, using eq. (3), we get the parameter set for $K=300$MeV and
$K_c=-3.990$MeV only at$M^*=0.83M$, in the case of the point-like nucleons
($R_n=0$)[11].
(We remark that the value of $\rho_0$ is slightly different from the one used
in this paper. However, the results hardly depend on $\rho_0$. )
In those cases, the number of the parameters may not be large enough to
reproduce the empirical value well, and, if the higher terms of $\phi$ are
added to (3), the wider range of $K$ and $M^*$ may be available.
However, $K=300\pm 50$MeV is $necessary$ to reproduce the empirical values of
$K$, $K_c$, and $K_{vs}$ simultaneously, for any type of $U(\phi )$ and for any
$R_n(=0\sim 0.8$fm).

(3) The results do not have a strong-dependence on $\rho_0$, $a_1$ and $a_4$,
since the calculated $K_{vs}$ is much more sensitive to the ratio $K'/K$ than
to those quantities.

In summary, we have studied $K$, $K_c$, and $K_{vs}$ by using the relativistic
mean field theories based on the $\sigma$-$\omega$-$\rho$ model with the
nonlinear $\sigma$ term and the excluded volume effects.
It is found that, at any $R_n(\leq 0.8$fm), $K=300\pm 50$MeV is $necessary$ to
account for the empirical values of $K$, $K_c$, and $K_{vs}$ at the same time,
as is in the cases of the point-like nucleons [17].
The result is independent on the detail descriptions of $U(\phi )$.
To see this result with the one in ref. [17], it seems that the conclusion that
$K=300\pm 50$MeV is $necessary$ to account for $K_{vs}$ is not drastically
changed, if we use any type of the relativistic mean-field theory and the
scaling model.
The reason is probably that the calculated $K_{vs}$ is most sensitive to the
ratio $K'/K$, which is adjusted to the empirical values.
This general conclusion $K=300\pm 50$MeV has also good agreement with the
result of the Dirac-Brueckner-Hartree-Fock calculation [19], with the result in
ref. [10], and with the earlier work by Sharma [20].

\bigskip

\noindent
$Acknowledgment$: One(H. K.) of the authors would like to thank Prof. M.M.
Sharma for useful discussions and suggestions. The authors gratefully
acknowledge the computing time granted by the Research Center for Nuclear
Physics (RCNP).



\bigskip

\bigskip

\centerline{{\bf{References}}}

\bigskip

\noindent
[1] J.P. Blaizot, Phys. Rep. {\bf{64}} (1980)171.

\noindent
[2] J.M. Pearson, Phys. Lett. B{\bf{271}} (1991)12.

\noindent
[3] S. Shlomo and D.H. Youngblood, Phys. Rev. {\bf{C47}} (1993)529.

\noindent
[4] S. Rudaz, P.J. Ellis, E.K. Heide and M. Prakash, Phys. Lett. {\bf{B285}}
(1992)183.

\noindent
[5] E.K. Heide and S. Rudaz, Phys. Lett. {\bf{B262}} (1991)375.

\noindent
[6] D. Von-Eiff, J.M. Pearson, W. Stocker and M.K. Weigel,

\noindent
Phys Lett. {\bf{B324}} (1994)279.

\noindent
[7] D. Von-Eiff, J.M. Pearson, W. Stocker and M.K. Weigel,

\noindent
Phys. Rev. {\bf{C50}} (1994)831.

\noindent
[8] D. Von-Eiff, W. Stocker and M.K. Weigel, Phys. Rev. {\bf{C50}} (1994)1436.

\noindent
[9] A.R. Bodmer and C.E. Price, Nucl. Phys. {\bf{A505}} (1989)123.

\noindent
[10] M.V. Stoitsov, P. Ring and M.M. Sharma, Phys. Rev. {\bf{C50}} (1994)1445.

\noindent
[11] H. Kouno, N. Kakuta, N. Noda, K. Koide, T. Mitsumori, A. Hasegawa and M.
Nakano, Phys. Rev. {\bf{C51}} (1995)1754.

\noindent
[12] H. Kouno, K. Koide, T. Mitsumori, N. Noda, A. Hasegawa and M. Nakano,
Phys. Rev. {\bf{C52}} (1995)135.

\noindent
[13] H. Kouno, K. Koide, T. Mitsumori, N. Noda, A. Hasegawa and M. Nakano,
preprint, SAGA-HE-87-95.

\noindent
[14] J. Boguta and A.R. Bodmer, Nucl. Phys. {\bf{A292}} (1977)413

\noindent
[15] A.R. Bodmer, Nucl. Phys. {\bf{A526}} (1991)703.

\noindent
[16] D.H. Rischke, M.I. Gorenstein, H. St{\"{o}}cker and W. Greiner,
Z. Phys. {\bf{C51}} (1991)485;

\noindent
Qi-Ren Zhang, Bo-Qiang Ma and W. Greiner, J. Phys. {\bf{G18}} (1992)2051;

\noindent
Bo-Qiang Ma, Qi-Ren Zhang, D.H. Rischke and W. Greiner, Phys. Lett. {\bf{B315}}
(1993)29.

\noindent
[17] H. Kouno, T. Mitsumori, N. Noda, K. Koide, A. Hasegawa and M. Nakano,
preprint, SAGA-HE-88-95.

\noindent
[18] B.D. Serot, Phys. Lett. B{\bf{86}} (1979)146:
%
B.D. Serot and J.D. Walecka, $The$ $Relativistic$ $Nuclear$ $Many$-$Body$
$Problem$ in: Advances in nuclear physics, vol. 16 (Plenum Press, New York,
1986).

\noindent
[19] R. Brockmann and R. Machleidt, Phys. Rev. {\bf{C42}} (1990)1965:
F. de Jong and R. Malfliet, Phys. Rev. {\bf{C44}} (1991)998:
H. Huber, F. Weber, and M.K. Weigel, Phys. Rev. {\bf{C51}} (1995)1790.

\noindent
[20] M.M. Sharma, preprint, DL/NUC/P323T.

\vfill\eject


\centerline{{\bf{Table and Figure Captions}}}

\bigskip

\noindent
Table 1

\noindent
The sets of the empirical values of $K$, $K_c$ and $K_{vs}$ in the table 3 in
ref. [2]. (According to the conclusion in ref. [2], we only show the data in
the cases of $K=150\sim 350$MeV.) All quantities in the table are shown in MeV.

\bigskip

\noindent
Table 2

\noindent
Range of the calculated $K_{vs}$ using the sets of $K$ and $K_c$ in table 1 as
inputs, if we put $M^*=0.5M\sim 0.93M$.
(a) The result in the case with excluded volume effects with $R_n=0.8$fm.
(b) The result in the case with excluded volume effects with $R_n=0.6$fm.
(c) The result in the case of the pointlike nucleons. ($R_n=0$)
:
In each table, "upper bound", "mean value", and "lower bound" mean that the
results are obtained by using the upper bound, the mean value, and the lower
bound of $K_c$ in table 1, respectively.
All quantities in the table are shown in MeV.

\bigskip

\noindent
Fig. 1~~~~~$K_{vs}$ as a function of $M^*$.
(a) The cases of $K=300$MeV and $K_c=-3.990$MeV.
(b) The cases of $K=350$MeV and $K_c=-7.724$MeV:
In each figure, the solid line is the result in the case of the point-like
nucleons, and the dotted and the dashed lines are the results in the cases with
excluded volue effects with $R_n=0.6$fm and with $R_n=0.8$fm, respectively.

\vfill\eject


\large

\hspace*{-2cm}
  \begin{tabular}{cccccc}
                                    \hline
    \     & Set 1 & Set 2 & Set 3 & Set 4 & Set 5  \\ \hline
    \   $K$ & 150.0 & 200.0  & 250.0 & 300.0 & 350.0 \\
    \  $K_c$ & $5.861\pm2.06 $ & $2.577\pm2.06 $ & $-0.7065\pm2.06 $
     & $-3.990\pm2.06 $ &$-7.274\pm2.06 $ \\
    \ $K_{vs}$ & $66.83\pm101$ & $-46.94\pm101$ & $-160.7\pm101$
     & $-274.5\pm101$ & $-388.3\pm101$ \\ \hline
  \end{tabular}

\bigskip

\begin{center}
Table 1
\end{center}

\bigskip

\bigskip

\bigskip


  \begin{tabular}{cccc}
                                    \hline
    \         &  upper bound   & mean value     & lower bound    \\ \hline
    \ $K=150$ & 1211$\sim$2562 &  960$\sim$2157 &  709$\sim$1752 \\
    \ $K=200$ &  808$\sim$1902 &  558$\sim$1497 &  307$\sim$1092 \\
    \ $K=250$ &  406$\sim$1242 &  155$\sim$ 836 &  -96$\sim$ 431 \\
    \ $K=300$ &    4$\sim$ 582 & -247$\sim$ 176 & -498$\sim$-229 \\
    \ $K=350$ & -399$\sim$ -79 & -652$\sim$-484 & -944$\sim$-889 \\ \hline
  \end{tabular}

\bigskip

\begin{center}
      Table 2(a)
\end{center}

\bigskip

\bigskip

\bigskip


  \begin{tabular}{cccc}
                                    \hline
    \         & upper bound    & mean value     & lower bound    \\ \hline
    \ $K=150$ & 1021$\sim$1962 &  795$\sim$1619 &  569$\sim$1276 \\
    \ $K=200$ &  659$\sim$1404 &  432$\sim$1062 &  206$\sim$ 719 \\
    \ $K=250$ &  296$\sim$ 847 &   70$\sim$ 504 & -156$\sim$ 161 \\
    \ $K=300$ &  -66$\sim$ 290 & -292$\sim$ -53 & -521$\sim$-396 \\
    \ $K=350$ & -428$\sim$-268 & -673$\sim$-610 & -959$\sim$-880 \\ \hline
  \end{tabular}

\bigskip

\begin{center}
      Table 2(b)
\end{center}

\bigskip

\bigskip

\bigskip


  \begin{tabular}{cccc}
                                    \hline
    \         & upper bound    & mean value     & lower bound    \\ \hline
    \ $K=150$ &  966$\sim$1755 &  749$\sim$1436 &  531$\sim$1118 \\
    \ $K=200$ &  618$\sim$1237 &  400$\sim$ 919 &  182$\sim$ 601 \\
    \ $K=250$ &  269$\sim$ 721 &   51$\sim$ 402 & -167$\sim$  84 \\
    \ $K=300$ &  -80$\sim$ 204 & -298$\sim$-115 & -522$\sim$-433 \\
    \ $K=350$ & -431$\sim$-313 & -671$\sim$-632 & -951$\sim$-865 \\ \hline
  \end{tabular}

\bigskip

\begin{center}
      Table 2(c)
\end{center}

\end{document}